\documentclass[referee]{raa}            

\usepackage[a4paper,margin=2.5cm]{geometry}
\usepackage{graphicx,times}            
\usepackage{natbib}
\usepackage{amssymb,amsmath}
\bibpunct{(}{)}{;}{a}{}{,}

\usepackage[pagebackref=true]{hyperref}

\begin{document}

  \title{Study of type-B QPOs observed in black hole X-ray binary Swift J1728.9-3613
}

   \volnopage{Vol.0 (20xx) No.0, 000--000}      
   \setcounter{page}{1}          

   \author{Raj Kumar 
   \inst{1,2}
   }

   \institute{Astrophysical Sciences Division, Bhabha Atomic Research Centre, Mumbai, 400085, India; {\it arya95raj@gmail.com}\\
        \and
             Homi Bhabha National Institute, Mumbai, 400094,  India\\
\vs\no
   {\small Received 20xx month day; accepted 20xx month day}}

\abstract{We report the detection of type-B quasi-periodic oscillation (QPO) of the black hole X-ray binary Swift J1728.9-3613 observed by \emph{NICER} during the 2019 outburst. A type-B QPO was observed for the first two days and it disappeared as flux increased, but again appeared at $\sim7.70$ Hz when flux was dramatically decreased. The source was found in the soft-intermediate state during these observations. We further studied the energy dependence of the QPO. We found that QPO was observed only for a higher energy range implying that the origin of QPO is possibly due to the corona emitting higher energy photons by the inverse Compton process. The variation of spectral parameters can be explained with the disk truncation model. The fractional rms found to be monotonically increased with energy. The phase lag spectrum followed the ``U-shaped'' curve. The rms and phase lag spectrum are modelled and explained with the single-component comptonization model \texttt{vkompthdk}.
}

   \authorrunning{Raj Kumar}
   \titlerunning{Type-B QPO : Swift J1728.9-3613}  

   \maketitle

%
%
\section{Introduction}           
\label{sect:intro}
\noindent Accretion disks formed around black hole X-ray binaries (BHXRBs) due to Roche lobe overflow or wind-fed accretion from a companion star to the black hole (BH), emits strong electromagnetic radiation in X-rays. BHXRBs are observed in two classes- (i) persistent X-ray sources and (ii) transient X-ray sources. Persistent sources are always observed in the bright phase. On the other hand, transient sources are observed in a bright phase known as an outburst, and then they go into a quiescent state for a long time, from months to decades. Based on the thermal emission and comptonized component, black hole transient evolves through different spectral states like low/hard state (LHS), high/soft state (HSS), hard intermediate state (HIMS), and soft intermediate state (SIMS) during its outburst \citep{Homan2005}. LHS is dominated by comptonized component with a photon index $\sim$ 1.7 and peak emission at $\sim$ 60-100 keV \citep{Joinet2008, Motta2009, Gilfanov2010}. In the LHS, the soft photons originating from the accretion disk are scattered by thermal electrons in the comptonization region. HSS is thermally dominated with disk temperature $\sim$ 0.5-1 keV \citep{Remillard2006} and a high energy tail can be produced by the scattering of soft photons by hybrid, thermal/nonthermal electrons \citep{Zdziarski2004}. Intermediate states are found between LHS and HSS but in general, HIMS is slightly harder than SIMS \citep{Belloni2011}.

\noindent As spectral behaviour changes with the evolution during an outburst, the timing properties also change for black hole transients \citep{Belloni2011}. The power density spectrum (PDS), is the most useful tool to probe the variability of X-ray binaries \citep{Belloni2005, Belloni2010, Marco2022}. In the study of XRBs, narrow peaks known as quasi-periodic oscillations (QPOs) are commonly observed in the X-ray PDS. QPOs are thought to originate from inner accretion flow but their exact origin is still debatable \citep{motta2011}. The low-frequency QPOs are classified into three categories: type-A, B and, C. Type-A QPOs are generally observed with a very weak, flat, and broad peak in the soft state in the frequency range of about 6-8 Hz \citep{Homan2005}. The type-B QPOs were observed in a narrow frequency range of 4-8 Hz with a weak read noise in PDS and fractional rms variability $\sim$ 5\% during SIMS \citep{Casella2005, Belloni2016}. Type-C QPOs are mostly observed during LHS and HIMS, with a frequency range 0.1-30 Hz with a flat top noise in PDS. Around 30\% fractional rms variability of Type-C QPOs was observed in LHS, which decreases to $\sim$10\% in HIMS \citep{Wijnands1999, Casella2005, Motta2016}.

\noindent Swift J1728.9-3613 is a newly detected X-ray transient discovered by the Burst Alert Telescope (BAT) onboard the SWIFT observatory \citep{Barthelmy2019}. \citet{Saha2023} studied the timing and spectral properties of Swift J1728.9-3613, identified it as a BHXRB, and put a lower limit on the BH mass $\sim$ 4.6 solar mass. \cite{Balakrishnan2023} concluded that the black hole in this system is likely associated with the supernova remnant G351.9-0.9. \cite{Draghis2023} measured the spin of black hole $\sim$0.86 using the relativistic reflection method and also measured the small inclination angle of the accretion disc $ i < 10^{\circ}$. \citet{Heiland2023} suggest that the viscous inflow time-scale of matter in the standard disc is responsible for the observed continuum lag of $8.4 \pm 1.9$ days between soft energy band $(2 - 4)$ keV and hard energy band $(10 - 20)$ keV. \citet{Heiland2023} also estimated the spin parameter of $\sim 0.6-0.7$ with an inclination angle of $\sim 45^{\circ} - 70^{\circ}$ using spectral analysis of \textit{NuSTAR} data. 

\noindent \emph{NICER} observed this source when it reached the SIMS, where type-B QPO is observed in the starting observation of the source \citep{Enoto2019}. In our work, we mainly focused on the study of Type-B QPO observed in Swift J1728.9-3613. Observations details are tabulated in Table \ref{tab1}. Observations and data analysis are discussed in section \ref{sect:Obs}. The results of our study are presented and discussed in section \ref{sect:Results}. We concluded our results in section \ref{sect:Conclusions}.

\section{Observations and Data analysis}
\label{sect:Obs}

\noindent \emph{NICER} is a payload on the International Space Station (ISS). It has an X-ray timing instrument (XTI) that operates in soft X-ray energy range (0.2-12 keV) \citep{Gendreau2016}. XTI is a collection of 56 X-ray concentrator optics with silicon drift detectors. Out of 56 detectors, 52 detectors are active at present. Here, we used 50 detectors for our analysis and discarded Focal Plane Modules (FPMs) 34 and 14 using the `detlist' flag because they are noisy under certain conditions. \emph{NICER} observed Swift J1728.9-3613 during the 2019 outburst. We mainly considered those observations which lie in SIMS. Details of observation used in our work are tabulated in Table \ref{tab1}. We used \texttt{HEASOFT-6.31.1} and \texttt{NICERDAS} $2022-12-16\_V010a$ for the analysis of the data. The \emph{NICER} CALDB version of 20221001 was used while extracting the level 2 data from level 1 data. Light curve was generated using the ``nicerl3-lc" pipeline with time bin 1 sec and 0.01 sec. We used the ``nicerl3-spec" pipeline to extract spectrum, arf, rmf, and background file. We added $1\%$ systematic error to the data and grouped each spectrum with at least 25 counts for each bin using the grppha task. We extracted the background for the \texttt{SCORPEON} model.

\begin{table}[h]
\centering
\caption{Details of \emph{NICER} observations used in this work.}\label{tab1}%
\begin{tabular}{llllll}
\hline
Obs ID     & Start time (MJD) & Stop time (MJD)  & \begin{tabular}[c]{@{}l@{}}average count rate\\ \emph{NICER} (1-10keV)\end{tabular} & QPO \\ \hline
1200550101 & 58512.64         & 58512.97        & 560              & yes \\
1200550102 & 58513.02         & 58513.68        & 666              & yes \\
1200550102 & 58513.72         & 58514.00        & 1128             & no  \\
1200550103 & 58514.03         & 58514.26        & 1316             & no  \\
1200550103 & 58514.36         & 58514.90        & 990              & yes \\ \hline
\end{tabular}
\end{table}

\begin{figure}[h]%
\centering
\includegraphics[width=0.7\textwidth, angle=0]{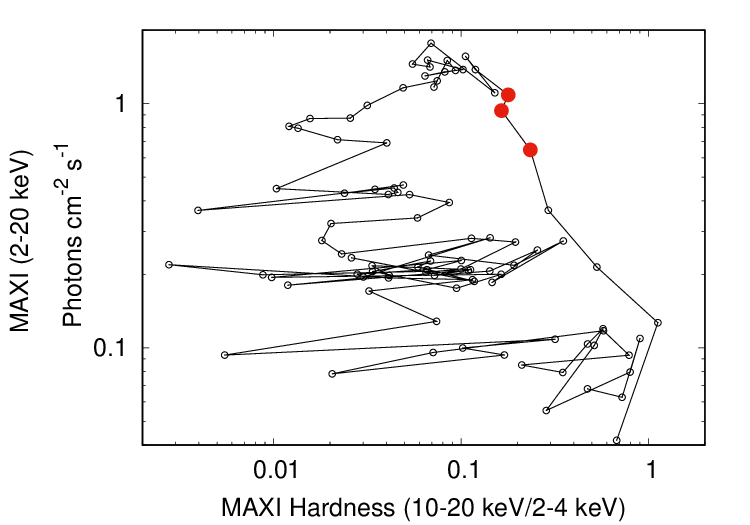}
\caption{Hardness intensity diagram of Swift J1728.9-3613 during 2019 outburst using MAXI/GSC. The hardness ratio was obtained from 2–4 keV and 10–20 keV photon flux. The red points indicate the NICER observations analyzed in this work.}\label{hardness}
\end{figure}

\begin{figure}[h]%
\centering
\includegraphics[width=0.7\textwidth, angle=270]{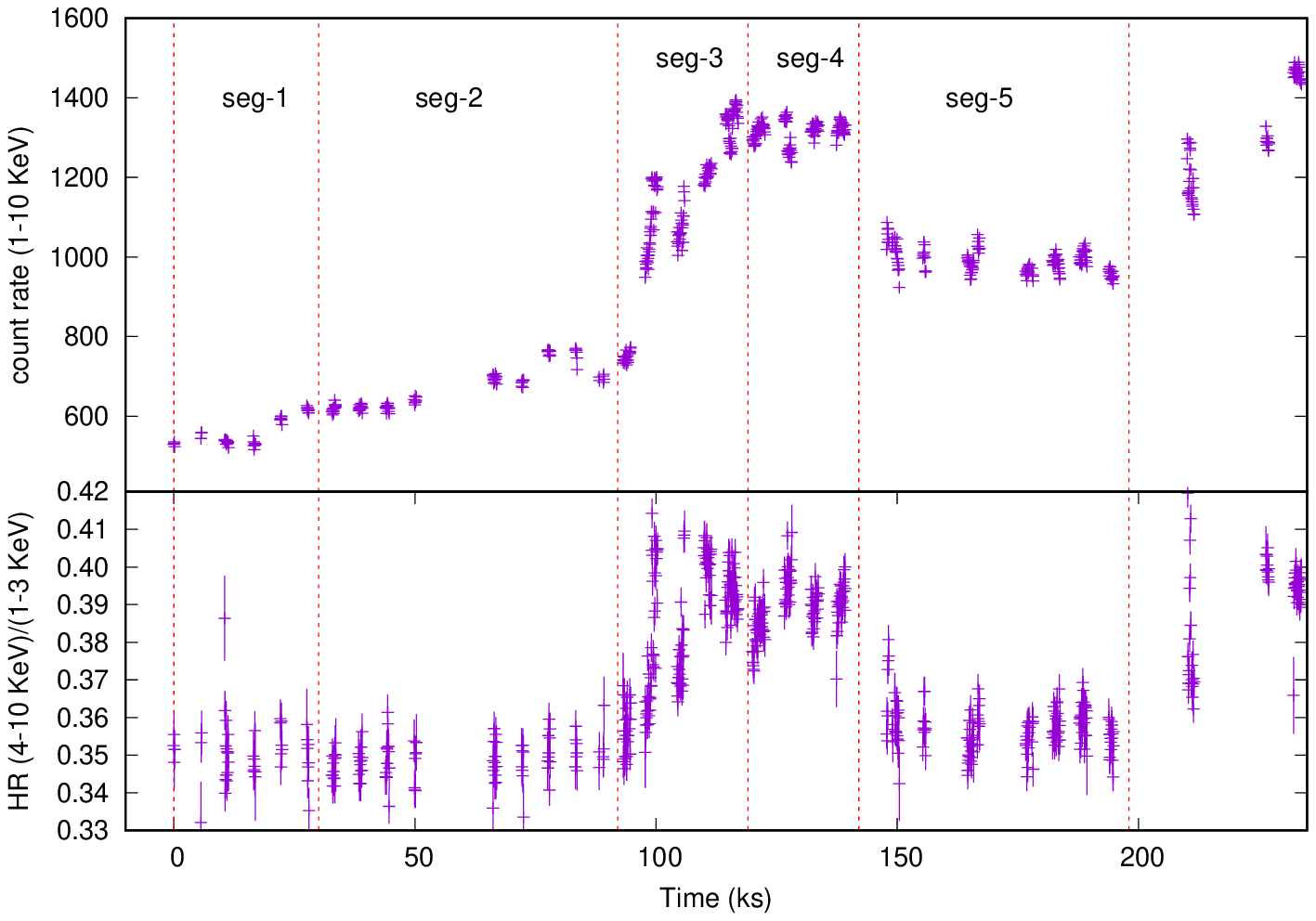}
\caption{\emph{NICER} lightcurve in 1-10 keV range and hardness ratio with bin size 60 sec. Hardness ratio (HR) is defined as the count rate ratio in 4-10 keV energy range to 1-3 keV energy range. Different segments are also shown in Figure.}\label{fig1}
\end{figure}
\noindent \citet{Saha2023} identified the Swift J1728.9-3613 as a black hole X-ray binary. They studied the evolution of outburst using hardness-intensity diagram (HID) and found that Swift J1728.9-3613 stayed in SIMS for $\sim$22 days. The HID of Swift J1728.9-3613 is shown in Figure \ref{hardness}. The observations used in this work are indicated by red points on HID. These observations belong to the SIMS as found by \citep{Enoto2019, Saha2023}. The upper panel of Figure \ref{fig1} shows the 1-10 keV \emph{NICER} lightcurve of Swift J1728.9-3613 from MJD 58512.6 to 58515.2 with bin size 60 sec. We divided the lightcurve into 5-segments. The count rate slowly increased from seg-1 to seg-4. Then count rate rapidly decreased by 25 $\%$ in seg-5. The average count rate in seg-5 reached 990 counts/sec from 1316 counts/sec in seg-4. We computed the hardness ratio (HR) by taking the ratio of the count rate in the energy range 4-10 keV and 1-3 keV. The HR increased in seg-3 and seg-4. Then it again decreased in seg-5. The HR is plotted in the lower panel of Figure \ref{fig1}.

\section{Results and Discussions}
\label{sect:Results}

\begin{table*}
\begin{center}
 
\hspace{0.25cm} {\caption{Fit parameters for power spectrum in energy range $1.0-10.0$ keV}\label{tab2}}
	\begin{tabular}{ccccccc}
 \hline
 \hline
 Model& Parameter & Seg-1 & Seg-2 & Seg-3 & Seg-4 & Seg-5 \\
	\hline
 Lorentzian 1 & $\nu$ (Hz) & 0(f) & 0(f)& 0(f) & - & 0(f) \\
    & $\Delta$ (Hz) & $2.23_{-0.54}^{+0.70}$ & $2.41_{-0.31}^{+0.37}$ & $0.31_{-0.18}^{+0.16}$  &-&$0.53_{-0.30}^{+0.29}$ \\
 & norm $(10^{-3})$ & $4.08_{-0.93}^{+0.95}$ & $3.93_{-0.38}^{+0.40}$ & $0.33_{-0.08}^{+0.09}$ &-&$0.58_{-0.32}^{+0.25}$ \\
  Lorentzian 2  & $\nu$ (Hz) & $2.96_{-0.24}^{+0.30}$ & $2.93_{-0.08}^{+0.07}$ & - & - & $3.57_{-0.36}^{+0.34}$ \\
& $\Delta$ (Hz) & $1.84_{-0.75}^{+1.56}$ & $0.71_{-0.23}^{+0.29}$ & - & - & $2.45_{-0.99}^{+1.45}$\\
 & norm $(10^{-3})$ & $2.17_{-0.87}^{+0.14}$ & $0.99_{-0.25}^{+0.28}$ & - & - & $0.78_{-0.42}^{+0.40}$\\
Lorentzian 3 & $\nu$ (Hz) & $5.46_{-0.06}^{+0.06}$ & $5.76_{-0.07}^{+0.07}$ & $8.03_{-0.86}^{+1.32}$ & - & $7.70_{-0.20}^{+0.19}$\\
 & $\Delta$ (Hz) & $0.76_{-0.22}^{+0.24}$ & $1.16_{-0.24}^{+0.28}$ & $2.23_{1.94}^{+4.01}$ & - & $1.84_{-0.55}^{+0.73}$\\
 & norm $(10^{-3})$ & $2.63_{-0.57}^{+0.50}$ & $2.00_{-0.31}^{+0.30}$ & $0.18_{-0.14}^{+0.23}$ & - & $0.64_{-0.16}^{+0.17}$\\
 Lorentzian 4  & $\nu$ (Hz) & $11.15_{-0.81}^{+0.72}$ & $10.83_{-0.89}^{+0.78}$ & - & - & $15.73_{-0.54}^{+0.39}$ \\
& $\Delta$ (Hz) & $2.23_{-1.46}^{+2.11}$ & $5.14_{-2.17}^{+3.74}$ & - & - & $1.15_{-1.09}^{+1.90}$ \\
 & norm $(10^{-3})$ & $1.24_{-0.47}^{+0.54}$ & $1.56_{-0.45}^{+0.61}$ & - & - & $0.18_{-0.09}^{+0.12}$ \\
  Lorentzian 5 & $\nu$ (Hz) & - & - & $1.40_{-1.40}^{+2.48}$ & - & $0.98_{-0.97}^{+0.32}$  \\
& $\Delta$ (Hz) & - & - & $6.61_{-1.59}^{+2.45}$ & - & $1.85_{-0.84}^{+2.21}$  \\
 & norm $(10^{-3})$ & - & - & $1.27_{-0.29}^{+0.32}$ & - & $0.83_{-0.42}^{+0.60}$ \\
   pow law & $\alpha$ & - & - & - & $0.96_{-0.20}^{+0.31}$ & - \\
& norm $(10^{-5})$ & - & - & - & $2.84_{-1.45}^{+1.36}$& -  \\
 \hline
 & chi2/dof & 52/57 & 48/57 & 85/60 & 61/60 &72/54\\
 \hline
\end{tabular}
\end{center}
\end{table*}

\subsection{Timing Analysis}
\noindent We extracted the lightcurve in the energy range 1-10 keV with a time resolution of 0.01 sec for each segment. Then we used the ``powspec" tool to calculate the power spectrum for an interval length of 10.24 s with Miyamoto normalization \citep{Miyamoto1991}. It corresponds to Nyquist frequency 50 Hz and 1024 $(2^{10})$ bins per interval. The PDS was rebinned using the logarithmic binning factor of 1.05. The PDS was converted to xspec readable file using the \texttt{ftflx2xsp} task. The PDS fitted with the combination of Lorentzians. The Lorentzian used in this work is given by, 
$$P(\nu) = \frac{r^2 \Delta}{2\pi}\frac{1}{\left [\left ( \nu - \nu_0 \right )^2 + \left ( \Delta/2 \right )^2  \right ]}$$

\begin{figure}[!h]
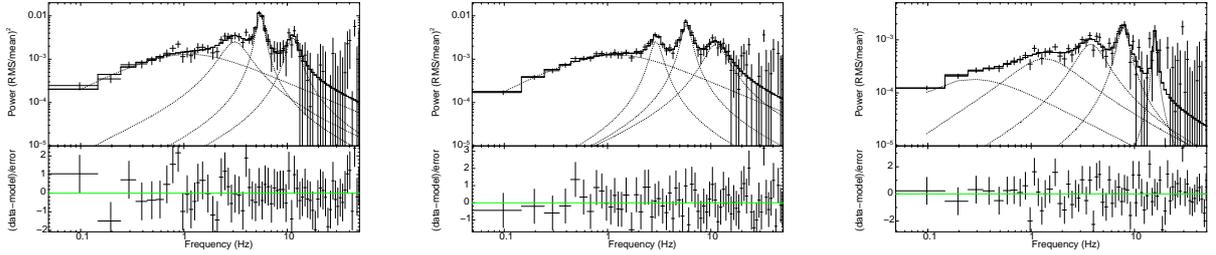
%
\centering
\subfloat {{\includegraphics[width=0.23\textwidth, angle=270]{ms2023-0271fig3.1.eps} }}%
\subfloat {{\includegraphics[width=0.23\textwidth, angle=270]{ms2023-0271fig3.2.eps} }}
\subfloat {{\includegraphics[width=0.23\textwidth, angle=270]{ms2023-0271fig3.3.eps} }}

\caption{Power density spectrum of Seg-1 (left panel), Seg-2 (middle panel) and Seg-5 (right panel). Type-B QPO was observed during these segments. The PDS of Seg-1 and Seg-2 were fitted with one zero-frequency Lorentzian and three Lorentzians. The PDS of Seg-5 was fitted with one zero-frequency Lorentzian and four Lorentzians. }\label{pds_spec}
\end{figure}

\noindent where $\nu_0$ is the centroid frequency, $\Delta$ is the full width at half maximum and $r$ is the integrated fractional rms \citep{Belloni2002}. In this definition, the quality factor Q is defined as $\sim \nu_0/\Delta$.  The significance of QPO is calculated by dividing the norm of power by 1 sigma negative error. Q value greater than 3 indicates the presence of QPO. The power spectra of Seg-1 and Seg-2 are fitted with a combination of 3 Lorentzians and one zero-frequency Lorentzian, while Seg-3 is fitted with a combination of 2 Lorentzians and one zero-frequency Lorentzian. Seg-4 is Poisson noise-dominated and fitted only with a power law, While we required 4 Lorentzian and one zero-frequency Lorentzian to fit the power spectra of Seg-5.

\noindent The PDS for each segment fitted with Lorentzians is shown in Figure \ref{pds_spec}. The details of fitted parameters are tabulated in Table 2. A LFQPO was observed in Seg-1 at $\sim 5.46$ Hz with Q value $\sim 7.2$ and rms $\sim 5.1 \% $ with a  significance of $\sim 8\sigma$. A harmonic was also observed at $\sim 11.15$ Hz with Q value $\sim 5.0$ and rms $\sim 3.5 \% $. The significance of harmonic is $\sim 4\sigma$. A LFQPO was observed in Seg-2 at $\sim 5.76$ Hz with Q value $\sim 5.0$ and rms $\sim 4.5\% $ with a significance of $\sim 10.8\sigma$. A subharmonic was also observed at $\sim 2.93$ Hz with Q value $\sim 4.1$ and rms $\sim 3.1 \% $. The significance of harmonic is $\sim 6.4\sigma$. A LFQPO was observed in Seg-5 at $\sim 7.7$ Hz with Q value $\sim 4.2$ and rms $\sim 2.5\% $ with a significance of $\sim 6.6\sigma$. A subharmonic and harmonic were also observed at $\sim 3.58$ Hz and $\sim 15.7$ Hz but the Q-value for subharmonic is less than 2 and the significance level for harmonic is less than $3.0\sigma$. There is no QPO was observed in Seg-3 and Seg-4.
\begin{figure}[!h]
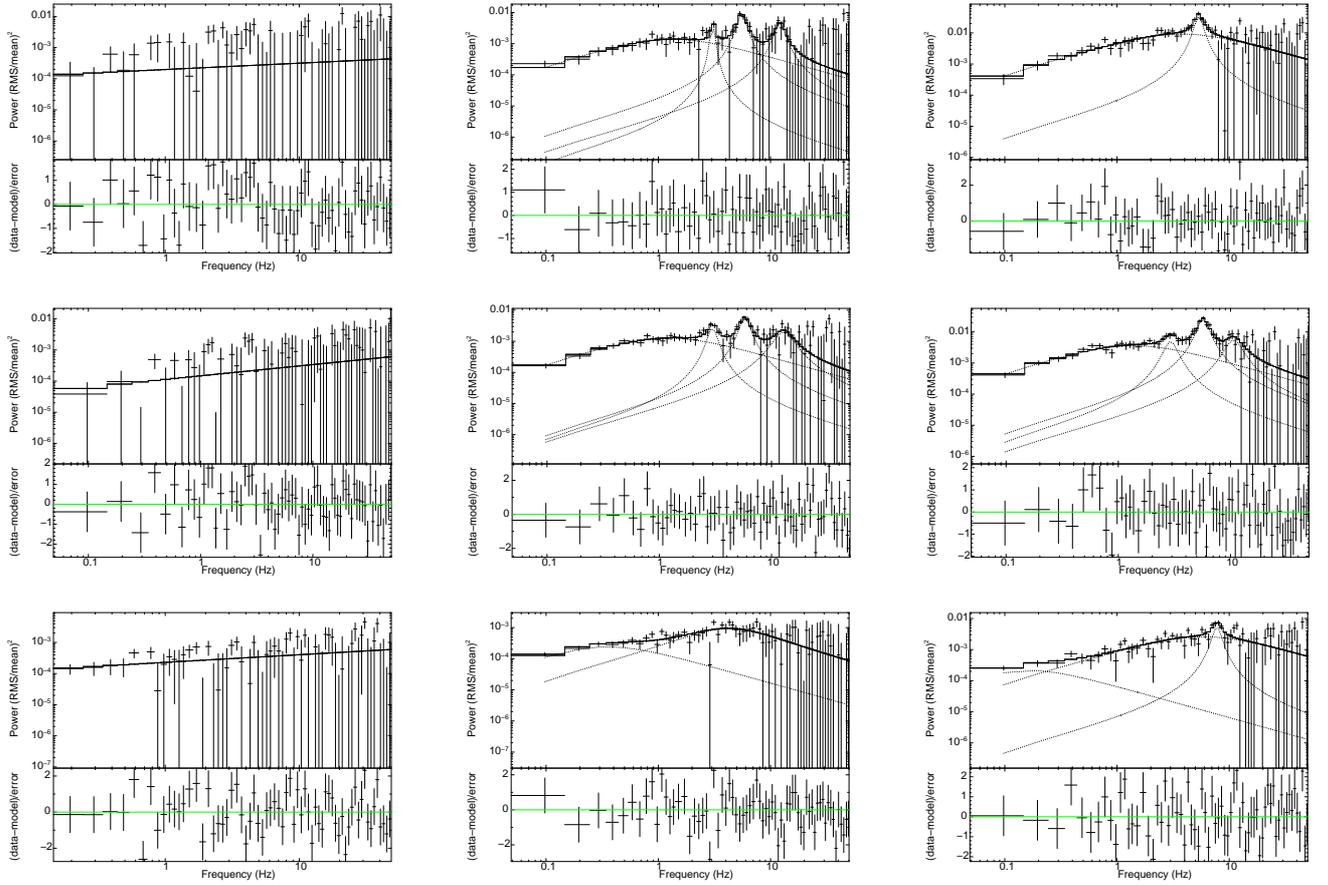
%
\centering
\subfloat {{\includegraphics[width=0.25\textwidth, angle=270]{ms2023-0271fig4.1.eps} }}%
\subfloat {{\includegraphics[width=0.25\textwidth, angle=270]{ms2023-0271fig4.2.eps} }}
\subfloat {{\includegraphics[width=0.25\textwidth, angle=270]{ms2023-0271fig4.3.eps} }}
\qquad
\subfloat {{\includegraphics[width=0.25\textwidth, angle=270]{ms2023-0271fig4.4.eps} }}%
\subfloat {{\includegraphics[width=0.25\textwidth, angle=270]{ms2023-0271fig4.5.eps} }}
\subfloat {{\includegraphics[width=0.25\textwidth, angle=270]{ms2023-0271fig4.6.eps} }}%
\qquad
\subfloat {{\includegraphics[width=0.25\textwidth, angle=270]{ms2023-0271fig4.7.eps} }}%
\subfloat {{\includegraphics[width=0.25\textwidth, angle=270]{ms2023-0271fig4.8.eps} }}
\subfloat {{\includegraphics[width=0.25\textwidth, angle=270]{ms2023-0271fig4.9.eps} }}%
\caption{The PDS of Seg-1 (upper panel), Seg-2 (middle panel) and Seg-5 (lower panel) in energy range 1-2 keV (left panel), 2-4 keV (middle panel) and 4-10 keV (right panel).}\label{pds_ene}
\end{figure}

\noindent To check the energy dependence of QPOs, we extracted the lightcurve in three different energy bands: 1-2 keV, 2-4 keV, and 4-10 keV. We calculated the PDS in each energy range for Seg-1, Seg-2 and Seg-5. Figure \ref{pds_ene} shows the PDS in energy ranges 1-2 keV, 2-4 keV, and 4-10 keV for Seg-1, Seg-2 and, Seg-5. In the soft energy band (1-2 keV), no QPO was observed for all segments. In energy band 2-4 keV, QPO was observed for Seg-1 and Seg-2 but no QPO was observed for Seg-5. In the energy band 4-10 keV, QPO was observed for all three segments. No signature of QPO in the soft energy band discards the origin of QPO from the disk. Therefore, the QPOs might be the result of oscillations in the corona. We will discuss the possible origin of QPOs in further detail in the following section.  

\subsection{Simultaneous fit of the photon, rms and phase-lag spectrum}
\begin{figure}[]
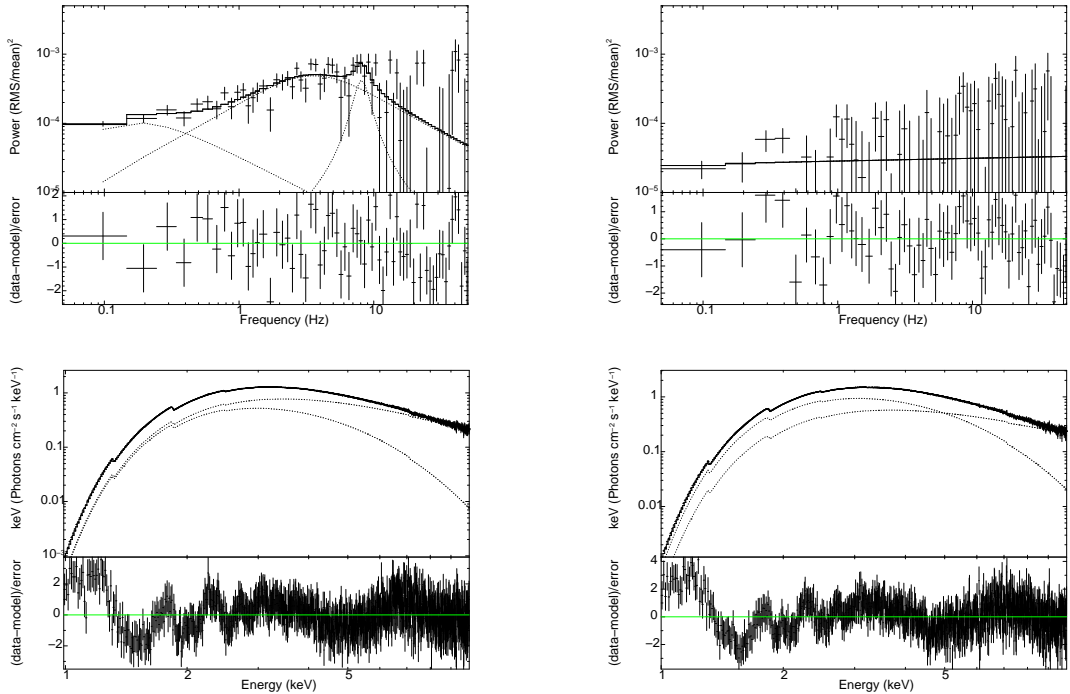
%
\centering
\subfloat {{\includegraphics[width=0.3\textwidth, angle=270]{ms2023-0271fig5.1.eps} }}%
\qquad
\subfloat {{\includegraphics[width=0.3\textwidth, angle=270]{ms2023-0271fig5.2.eps} }}
\qquad
\subfloat {{\includegraphics[width=0.3\textwidth, angle=270]{ms2023-0271fig5.3.eps}} }%
\qquad
\subfloat {{\includegraphics[width=0.3\textwidth, angle=270]{ms2023-0271fig5.4.eps}} }%
\caption{The PDS and energy spectrum of Seg-4 (left panel) and Seg-5 (Right panel). The type-B QPO was not observed during these segments.}\label{noqpo}
\end{figure}

\begin{figure}[]%
\centering
\subfloat {{\includegraphics[width=0.6\textwidth, angle=270]{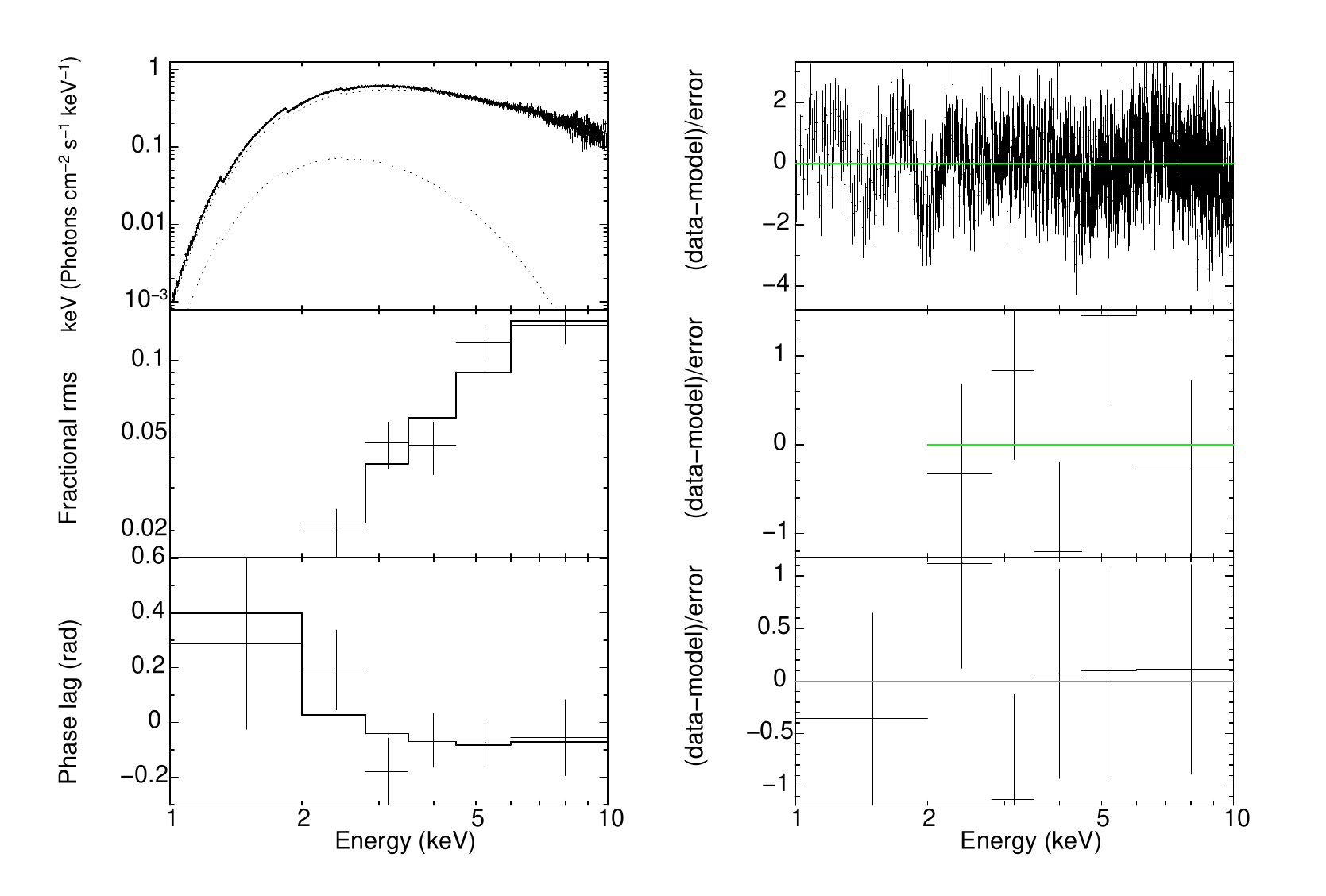}} }%
\caption{Simultaneous fitting of the photon, Fractional rms, and phase lag spectra for Seg-1. The rms and phase lag calculated at $\sim5.46$ QPO frequency.}\label{fig_vkompthdk1}
\end{figure}

\begin{figure}[h]%
\centering
\subfloat {{\includegraphics[width=0.6\textwidth, angle=270]{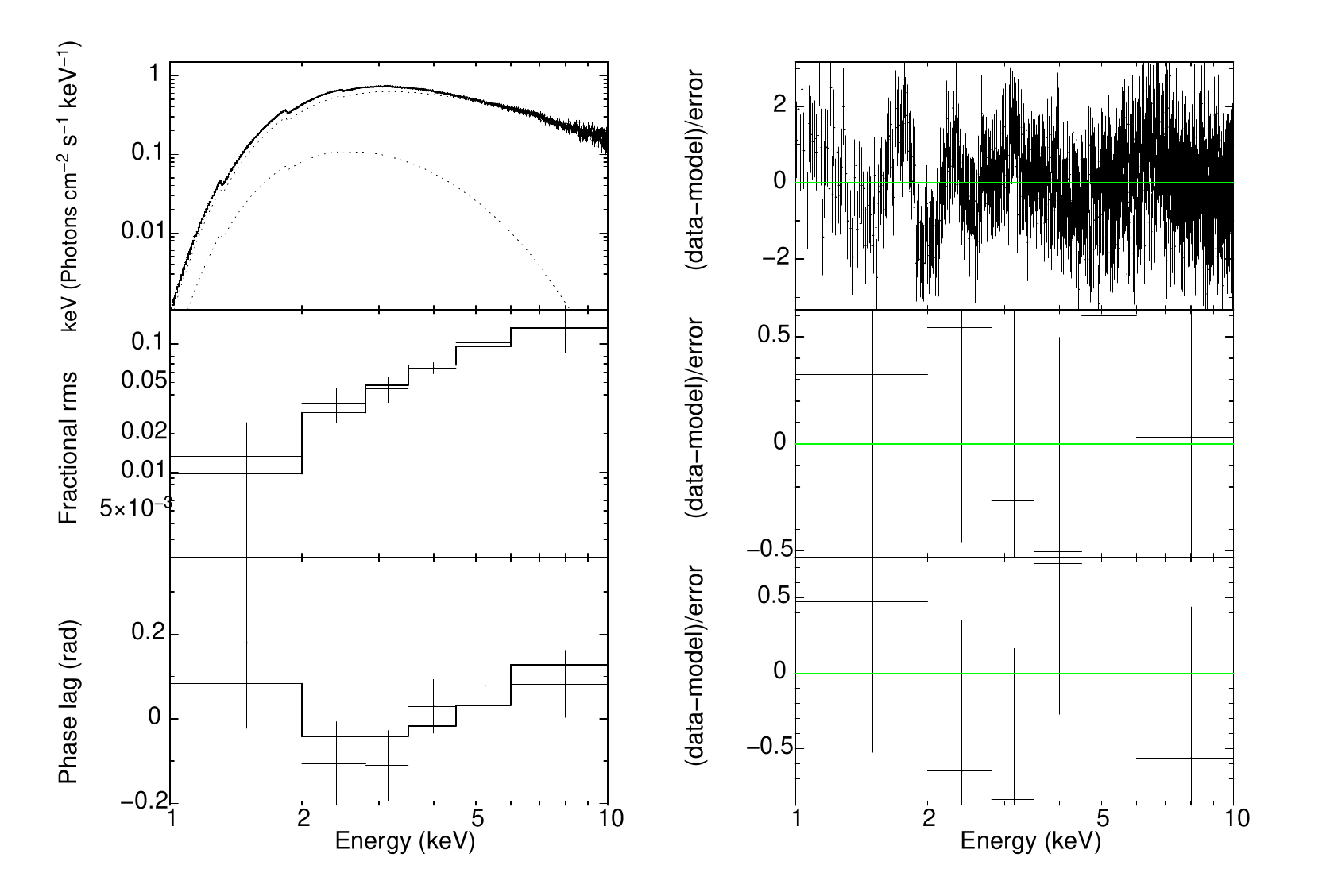} }}%
\caption{Simultaneous fitting of the photon, Fractional rms, and phase lag spectra for Seg-2. The rms and phase lag calculated at $\sim5.76$ Hz QPO frequency.}\label{fig_vkompthdk2}
\end{figure}

\begin{figure}[h]%
\centering
\subfloat {{\includegraphics[width=0.6\textwidth, angle=270]{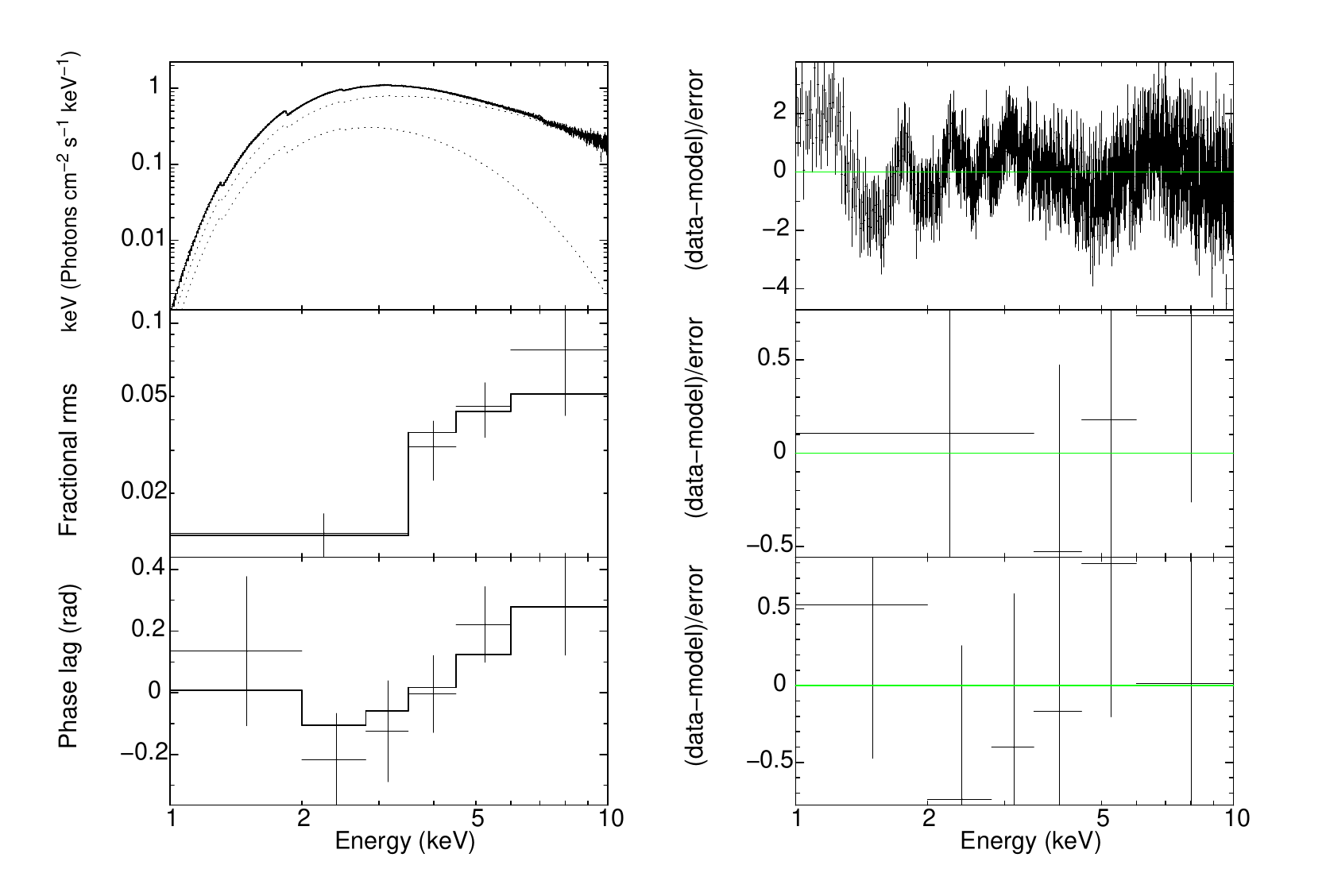} }}%
\caption{Simultaneous fitting of energy spectra, Fractional rms, and phase lag spectra for Seg-5. The rms and phase lag calculated at $\sim7.70$ Hz QPO frequency.}\label{fig_vkompthdk5}
\end{figure}

\noindent To move further and to extract a deeper understanding of the system, we estimated the frequency-dependent phase lag in the energy ranges 1.0-2.0 keV, 2.0-2.8 keV, 2.8-3.5 keV, 3.5-4.5 keV, 4.5-6.0 keV and 6.0-10.0 keV. The phase lag is calculated with reference energy band 1.0-10.0 keV. $\phi(\nu)$ is the phase of the cross-spectrum as a function of frequency (e.g. \citet{Nowak1999}) and it is related to the time lags between the selected bands as $\tau(\nu) = \phi(\nu)/2\pi\nu$. The phase lag is calculated in the frequency range of FWHM/2 at QPO frequency. We further used these energy ranges to calculate the rms spectra at observed QPO frequencies. To calculate the rms, we fitted the power spectra in each energy range by fixing the frequency of Lorentzian as given in Table \ref{tab2} for Seg-1 and Seg-2. For Seg-5, we calculated the rms in the energy range 1.0-3.5 keV, 3.5-4.5 keV, 4.5-6.0 keV and 6.0-10.0 keV. The error in rms calculated at 1-sigma level. The fractional rms at QPO frequency in each segment increased monotonically with energy. \cite{bendat2011random} derived the error formula for a single power spectrum and a single cross-spectrum. \citet{Ingram2019} presented new error formulae for the rms spectrum and energy-dependent cross-spectrum to overcome the issue of over-fitting due to formulae presented in \citet{bendat2011random}. In this work, we used formulae derived by \citet{Ingram2019} to calculate rms spectrum and energy-dependent time/phase lag spectrum. The definition used by \citet{Nowak1999} and \citet{Ingram2019} is that +ve lag corresponds to hard lags soft. So, In the current scenario, the +ve lag corresponds to the subject band lags reference band. For the number of realizations, N $>500$, We used a bias term equal to zero as recommended by \citet{Ingram2019}. Here N is the product of the number of segments in observation with the number of frequencies binned in the range. 
\noindent We simultaneously fitted the photon, rms and phase-lag spectra for Seg-1, Seg-2 and Seg-5 data sets while we fitted only the photon spectra for Seg-3 and Seg-4 using the model \texttt{phabs*(diskbb+nthcomp)+dilution*vkompthdk} within the software package XSPEC V12.13.0c. The dilution parameters is defined as$$dilution=\frac{flux(nthcomp)}{flux(diskbb+nthcomp)}$$
\noindent In this model, the soft photons are supplied to the corona from the accretion disk (\texttt{diskbb}). Soft photons are inverse-Compton scattered by the hot electrons in the corona (\texttt{nthcomp}). The scattered photons finally escape the corona with higher energy forming the spectrum of the source. In this model, QPO is considered as a small oscillation of the source spectrum around the time-averaged spectrum. This model can generate both hard and soft lag in the time variation of the observed photon flux. \texttt{Phabs} describes the photo-electric absorption of X-rays. Even though \texttt{vkompth} has the provision to take multiple coronae, in our analysis we used only one corona. We fixed $kTe$ of the \texttt{nthcomp} at 50 keV. An edge at $\sim 1.85$ keV for Seg-3, Seg-4 and Seg-5 was used to account for the residuals observed possibly due to instrumental effects. The \texttt{vkompth} model contains the seed photon temperature, the electron temperature, the power law photon index, the size of the corona, the feedback fraction and the variation of the external heating rate. In simultaneous fitting, a \texttt{dilution} component was used for the rms spectrum to take care of the dilution effects due to disk. we tied the kTe and gamma of \texttt{vkompthdk} with KTe and gamma of \texttt{nthcomp}. The fitted parameters are given in the Table \ref{tab3}. Type-B QPO was not observed for Seg-3 and Seg-4. The PDS and energy spectrum for Seg-3 and Seg-4 are shown in Figure \ref{noqpo}. Type-B QPO was observed for Seg-1, Seg-2 and Seg-5. For Seg-1, We obtained a corona size of $\sim1705$ km, with the feedback fraction of $10\%$. The seed photon source temperature of the corona was found to be $kT_s=0.861_{-0.159}^{+0.853}$ keV. This is higher than the temperature of the inner disc, $kT_{in}=0.72\pm0.02$. The variation of the external parameter found to be $\delta \dot{H}_{ext}=0.25_{-0.07}^{+0.32}$. For Seg-2, We obtained a corona size of $\sim3926$ km, with the feedback fraction of $9.4\%$. The seed photon source temperature of the corona was found to be $kT_s=0.517_{-0.137}^{+0.138}$ keV. This is lower than the temperature of the inner disc, $kT_{in}=0.77\pm0.02$. The variation of the external parameter found to be $\delta \dot{H}_{ext}=0.15_{-0.04}^{+0.07}$. For Seg-5, We obtained a corona size of $\sim7739$ km, with the feedback fraction of $56\%$. The seed photon source temperature of the corona was found to be $kT_s=0.332_{-0.105}^{+0.133}$ keV. This is lower than the temperature of the inner disc, $kT_{in}=0.92\pm0.01$. The variation of the external parameter found to be $\delta \dot{H}_{ext}=0.075_{-0.034}^{+0.095}$. The simultaneous fitting of the photon, rms and lag spectra for Seg-1, Seg-2, and Seg-5 are shown in Figure \ref{fig_vkompthdk1}, Figure \ref{fig_vkompthdk2} and Figure \ref{fig_vkompthdk5} respectively. \\

\noindent Over the years, an enormous amount of research has been carried out to understand the behaviour of the accretion process around astrophysical objects. The truncation disk model is a widely used model to explain the accretion process in moderately low accretion black hole binary systems. In the canonical truncation disk, the outer region is the standard thin disk \citep{Shakura1973} and the inner region consists of radiatively inefficient hot gas \citep{Zdziarski2004, McClintock2006, Done2007}. The disk contribution to the total intrinsic photon flux was increased from Seg-1 to Seg-4 and then decreased in Seg-5. The QPO was observed when the disk contribution was low or the corona contribution was more in the intrinsic photon flux. \\

\begin{table}[]
\centering
\hspace{0.25cm} {\caption{The Simultaneous fitting of the photon, rms and phase lag parameters for Seg-1, Seg-2, and Seg-5. For Seg-3 and Seg-4, the spectral fitting parameters are given. Here, $n_H$: equivalent neutral
hydrogen column density, $kT_{in}$: inner disk temperature, $N_{dbb}$: normalization of multicolor disk model \texttt{diskbb}, $\Gamma$:  asymptotic power-law index of the Comptonized photon distribution, norm: normalization of \texttt{nthcomp} model, $kT_s$: the seed photon temperature of \texttt{vkompthdk}, $L$: the size of the corona, $\eta$: the feed fraction, $\delta \dot{H}_{ext}$: the variation of the external heating rate. The confidence ranges for the error bars are $68\%$ CI. $F_{Total}$: the total unabsorbed photon flux in the energy range 1.0–10.0 keV, $F_{diskbb}$: the photon flux from a thin accretion disc around a black hole, and $F_{nthcomp}$: the photon flux due to the comptonized component. The units of flux is $erg\;cm^{-2}\;s^{-1}$.}\label{tab3}}
\begin{tabular}{lllllll}
\hline
Component & Parameter & Seg-1 & Seg-2 & Seg-3 & Seg-4 & Seg-5 \\ \hline 
\texttt{Phabs}     & $N_H (\times10^{22} cm^{-2})$      &   $3.14_{-0.01}^{+0.01}$    &     $3.13_{-0.01}^{+0.01}$  &  $3.24_{-0.01}^{+0.01}$    & $3.28_{-0.01}^{+0.01}$ & $3.21_{-0.01}^{+0.01}$   \\
\texttt{diskbb}    & $kT_{in}$      &   $0.72_{-0.02}^{+0.02}$    &   $0.77_{-0.02}^{+0.02}$     &    $1.04_{-0.01}^{+0.01}$&    $1.11_{-0.01}^{+0.01}$     &    $0.92_{-0.01}^{+0.01}$    \\
          & $N_{dbb}$      &  $260_{-42}^{+34}$     &    $267_{-27}^{+23}$     &    $343_{-14}^{+13}$&    $465_{-8}^{+8}$&    $367_{-15}^{+14}$     \\
\texttt{nthcomp}   & $\Gamma$     &   $2.54_{-0.03}^{+0.03}$    &   $2.61_{-0.03}^{+0.02}$     &    $2.82_{-0.06}^{+0.06}$&    $2.46_{-0.09}^{+0.09}$    &    $2.80_{-0.03}^{+0.03}$    \\
          & norm      &   $1.62_{-0.11}^{+0.12}$    &    $1.75_{-0.09}^{+0.10}$    &    $1.70_{-0.12}^{+0.12}$ &   $1.10_{-0.11}^{+0.11}$ &    $1.92_{-0.10}^{+0.11}$    \\
\texttt{vkompthdk} & $KTs(keV)$  &  $0.861_{-0.159}^{+0.853}$     &    $0.517_{-0.137}^{+0.138}$    & - & -  &   $0.332_{-0.105}^{+0.133}$     \\
          & L (km)    &   $1705_{-1310}^{+3402}$     &    $3926_{-2368}^{+4347}$   & - & - &   $7739_{-4438}^{+8437}$    \\
          & $\eta$        &  $0.10_{-0.06}^{+0.06}$     &   $0.094_{-0.075}^{+0.155}$    & - & - &   $0.55_{-0.42}^{*}$    \\
          & $\delta \dot{H}_{ext}$ &  $0.25_{-0.07}^{+0.32}$     &   $0.15_{-0.04}^{+0.07}$    & - & - &  $0.075_{-0.034}^{+0.095}$     \\ \\\hline
          &$F_{Total}(\times10^{-9})$& $9.09\pm0.01$ &$10.54\pm0.01$ &$17.54\pm0.02$ &$20.16\pm0.01$ &$15.21\pm0.01$ \\ 
          &$F_{diskbb}(\times10^{-9})$& $0.90\pm0.01$ &$1.27\pm0.01$ &$6.44\pm0.02$&$11.65\pm0.01$ &$3.99\pm0.01$ \\ 
          &$F_{nthcomp}(\times10^{-9})$& $8.19\pm0.01$ &$9.27\pm0.01$ &$11.10\pm0.02$ &$8.51\pm0.01$ &$11.22\pm0.01$ \\
          &$\frac{F_{diskbb}}{F_{Total}}$& $0.10$ &$0.12$ &$0.37$ &$0.58$ &$0.26$ \\\\\hline
          & $\chi^2/dof$  &  940/867     &    960/901   &    970/893 &    1145/893 &    967/897   \\ \hline
\end{tabular}
\end{table}
 
\noindent The fundamental physical concept related to the appearance of SIMS and the associated type-B QPO is still unresolved. Using RXTE data, \citet{Van2017} and \citet{Gao2017} noted that the lags of type-B QPOs are hard for low-inclination BHB systems, while lags are either hard or soft for high inclination BHBs. In the literature on lags, there seems to be general agreement that inverse-Compton scattering of the soft photons in corona produces hard lags \citep{Kylafis2008, Miyamoto1988} while down-scattering of the hard photons in the disk produce the soft lags \citep{Uttley2014}. \citet{Belloni2020} provided one of the earliest discussions of the positive lag below 2 keV (0.7-2.0 keV) energy range. Their study of type-B QPO in MAXI J1348-630 found that the phase lags in the 3-10 keV and the 0.7-2.0 keV are positive with respect to reference band 2-3 keV. In their study, they discarded the possibility of the origin of type-B QPOs due to the propagation of mass accretion rate fluctuations. \citet{Belloni2020} demonstrated that comptonization of flat seed photon spectrum between 2 and 3 keV with no emission at other energies can explain the positive lags of both low and high energies. This behaviour will change for the more realistic seed photons \citep{Peirano2023, Zhang2023}. The comptonization model \texttt{vkompth} \citep{Karpouzas2020, Bellavita2022} can account for the positive time lag at low energies and high energies. In Swift J1728.9-3613, we observed a ``U-shaped" curve in the Phase lag spectrum. This behaviour indicates that the origin of type-B QPO in Swift J1728.9-3613 is related to the corona. 

\noindent The \texttt{vkompth} has two different types based on the corona. \texttt{vkompthdk} is a model for one corona, where the corona is considered spherically symmetric. \texttt{vkdualdk} is model for two coronae. It is considered a small corona and a large corona. Small corona are located near the black hole and large corona can extend horizontally over the inner part of the accretion disk or extend vertically. This model is applied to investigate the geometry and disc-corona coupling. This model is used to explain the rms and phase lag of type-A QPO in MAXI J1348-630 \citep{ZhangL2023}, type-B QPO in MAXI J1348-630 \citep{Garcia2021,Bellavita2022}, GX 339-4 \citep{Peirano2023}, MAXI J1535-571 \citep{ZhangY2023}, MAXI J1820+070 \citep{Ma2023}, GRO 1655-40 \citep{Rout2023} and type-C QPO in MAXI J1820+070 \citep{Ma2023}, GRS 1915+105 \citep{Karpouzas2021,Garcia2022,Mendez2022}, MAXI J1535-571 \citep{Rawat2023}, GRO J1655-40 \citep{Rout2023}. For the type-B QPOs in Swift J1728.9-3613, we explained the rms and phase lag by one corona with varying size.


\section{Summary and Conclusions}
\label{sect:Conclusions}
\noindent In a nutshell, the \emph{NICER} observed the BHXRB Swift J1728.9-3613 when the source was in the soft intermediate state. A type-B QPO was observed during MJD 58512. During MJD 58513 the flux and hardness increased and type-B QPO disappeared. On MJD 58514, it showed a rapid transition to SIMS with the presence of type-B QPO. \citet{Xu2019} observed a rapid flux decrease $\sim 45\%$ in $\sim 40 s$ with turn-on of a QPO in BHXRB Swift J1658.2-4242 with NUSTAR, Swift and XMM-Newton. They suggest that the accretion disk instabilities triggered at a large disk radius cause the fast transition in spectral and timing properties of Swift J1658.2-4242. It is to be noted that the type-B QPO was not observed in the soft energy band (1-2 keV) for any segment but it was present in the hard energy band 2-10 keV for seg-1, seg-2 and 4-10 keV for seg-5. The photons in 2-10 keV in the SIMS of BHXRB are produced by thermal comptonization of the disc photons by the thermal electron present in the corona. Therefore, the presence of QPO in 2-10 keV / 4-10 keV possibly indicates that the QPO originated due to oscillations in comptonizing corona. The rms was found to be steadily increasing with energy for observed QPO. \\
\noindent In this work, we presented the study of spectral and timing properties of Swift J1728.9-3613 for MJD 58512, 58513 and 58514 using \emph{NICER} observations. During the analyzed time, the source was in SIMS. The primary results are:-
\begin{enumerate}
\item We reported a type-B QPO at $\nu\sim 7.70$ Hz. This QPO was observed when there was a rapid decrease in the count rate from 1316 cts/sec to 990 cts/sec during MJD 58514.
\item Energy-dependent PDS reveals that observed QPO is not present in the low energy band i.e. 1-2 keV band but it starts to appear at a higher energy range (after 2 keV for seg-1 and seg-2 but after 4 keV for seg-5). The rms and phase lag spectrum also indicate that observed QPO is modulated in the corona.
\item The rms and phase lag spectra are explained with the spherically symmetric corona. The size of the corona increased and the seed photon source temperature of the corona decreased with an increase in QPO frequency. 
\end{enumerate}

\begin{acknowledgements}
\noindent This work has made use of public  \emph{NICER}  data available at \url{https://heasarc.gsfc.nasa.gov/FTP/nicer/data/obs/} and the software provided by the High Energy  Astrophysics Science Archive Research Center (HEASARC) available at \url{https://heasarc.gsfc.nasa.gov/docs/software/lheasoft/}. The time-dependent Comptonization model is available at the GitHub repositories \url{https://github.com/candebellavita/vkompth}. The author thanks Dr. Adam Ingram for help in calculating the time lag, prof. Mariano Mendez for help in \texttt{vkompthdk} model and the anonymous reviewer for their useful comments which improved the quality of the work. 
\end{acknowledgements}

\bibliography{bibtex}{}
\bibliographystyle{raa}

\label{lastpage}

\end{document}